\documentclass[11pt]{article}
\usepackage[utf8]{inputenc}
\usepackage{mathrsfs}  
\usepackage{titlesec}

\titleformat*{\section}{\large\bfseries}
\usepackage{setspace}
\usepackage{amssymb,amsmath,amsthm,amsfonts,amstext}
\usepackage[english]{babel}
\usepackage[left=1.5cm,right=1.5cm,top=1.5cm,bottom=1.5cm]{geometry}
\usepackage{enumerate}
\usepackage[mathscr]{euscript}

\usepackage{mathtools,braket}
\usepackage{bm, hyperref}
\usepackage{xcolor, float}
\usepackage[capitalize]{cleveref}

\newtheorem{theorem}{Theorem}

\newtheorem{proposition}[theorem]{Proposition}
\newtheorem{lemma}[theorem]{Lemma}

\title{\Large
	\textbf{Extensions of the $(p,q)$-Flexible-Graph-Connectivity model}
}

\author{\large
Ishan Bansal\thanks{
	{\tt ib332@cornell.edu}. Cornell University, Ithaca, NY, USA.}
\and
Joseph Cheriyan\thanks{
{\tt jcheriyan@uwaterloo.ca}.
	Department of Combinatorics and Optimization, University of Waterloo, Canada.}
\and
Logan Grout\thanks{
	{\tt lcg58@cornell.edu}.
	Cornell University, Ithaca, NY, USA.}
\and 
\and
Sharat Ibrahimpur\thanks{
	{\tt s.ibrahimpur@lse.ac.uk}.
	London School of Economics and Political Science, London, UK.}
}

\date{}

\usepackage{tikz}
\usepackage{hyperref}

\usepackage[mathscr]{euscript}

	\newcommand{\Rp}{\ensuremath{\mathbb R_{\geq 0}}}
	
	\newcommand{\Zintp}{\ensuremath{\mathbb Z_{\geq 0}}}

	\newcommand{\opt}{\textsc{opt}}
	
	\newcommand{\safe}{\mathscr{S}}
	\newcommand{\unsafe}{\mathscr{U}}

	\newcommand{\fgc}{\mathrm{FGC}}
	\newcommand{\pqfgc}{(p,q)\text{-}\fgc}

	\newcommand{\genfgc}[2]{\ensuremath{(#1,#2)}\text{-}\fgc}
	\newcommand{\capndp}{\mathrm{Cap\text{-}NDP}}

	\newcommand{\sndp}{\mathrm{SNDP}}
	\newcommand{\NCsndp}{\mathrm{NC}\text{-}\mathrm{SNDP}}
	\newcommand{\pNCsndp}{p\text{-}\mathrm{NC}\text{-}\mathrm{SNDP}}
	\newcommand{\NCfgc}{\mathrm{NC}\text{-}\fgc}
	\newcommand{\pNCfgc}{p\text{-}\mathrm{NC}\text{-}\fgc}

	\newcommand{\fst}{\mathrm{FST}}
	\newcommand{\term}{\ensuremath{T}}

	\newcommand{\twoecs}{\mathrm{Steiner}\text{-}\mathrm{2ECS}}

\newcommand\dlambda{\overrightarrow{\lambda}}
\newcommand\dH{\overrightarrow{H}}

\begin{document}

\maketitle

\begin{abstract}
{
We present approximation algorithms for network design
problems in some models related to the $\pqfgc$ model.
Adjiashvili, Hommelsheim and M\"uhlenthaler \cite{AHM21,AHM20}
introduced the model of Flexible Graph Connectivity that we denote by $\fgc$.
Boyd, Cheriyan, Haddadan and Ibrahimpur \cite{BCHI21} introduced a generalization of $\fgc$.
Let $p\geq 1$ and $q\geq 0$ be integers.
In an instance of the $(p,q)$-Flexible Graph Connectivity problem, denoted $\pqfgc$,
we have an undirected connected graph $G = (V,E)$, a partition
of $E$ into a set of safe edges $\safe$ and a set of unsafe edges
$\unsafe$, and nonnegative costs $c\in\Rp^E$ on the edges.
A subset $F \subseteq E$ of edges is feasible for the $\pqfgc$ problem if
for any set $F'\subseteq\unsafe$ with $|F'|\leq q$,
the subgraph $(V, F \setminus F')$ is $p$-edge connected.
The algorithmic goal is to find a feasible edge-set $F$ that
minimizes $c(F) = \sum_{e \in F} c_e$.

We introduce a generalization of the $\genfgc{p}{q}$ model, called
$\genfgc{\{0,1,\dots,p\}}{\{0,\dots,q\}}$,
where a required level of edge-connectivity $p_{ij} \in \{0,\dots,p\}$
and a fault-tolerance level $q_{ij}\in \{0,\dots,q\}$ is specified
for every pair of nodes $\{i,j\}$.
The goal is to find a subgraph $H=(V,F)$ of minimum cost such that
for any pair of nodes $\{i,j\}$ and any set of at most $q_{ij}$
unsafe edges $F'\subseteq F$, the graph $H - F'$ has $p_{ij}$
edge-disjoint $(i,j)$-paths.
Assuming that $p=1$ or $q=1$ (i.e., when $p_{ij}\in\{0,1\}$ or when $q_{ij}\in\{0,1\}$),
we present $\max(2(p+1),\;2(q+1))$-approximation algorithms for this model
by reductions to the capacitated network design problem ($\capndp$);
we apply Jain's iterative rounding method \cite{Jain01} to solve the $\capndp$~instances.

We also consider the Flexible Steiner Tree model, denoted $\fst$.
We present a straight-forward approximation algorithm for $\fst$
that achieves approximation ratio~$\approx{2.9}$ via recent results
of Ravi, Zhang \& Zlatin \cite{RZZ22}.

Finally, we introduce the $\NCfgc$ model.
In an instance of this problem, we have an undirected connected graph
$G = (V,E)$, a partition of $V$ into a set of safe nodes $V^S$ and a
set of unsafe nodes $V^U$, and non-negative costs $c\in\Rp^E$ on the
edges; moreover, for every pair of nodes $\{s,t\}$, a required level of
connectivity $r_{st}\in\Zintp$ is specified.
The goal is to find a subgraph $H=(V,F)$ of minimum cost such that for
any pair of nodes $\{s,t\}$, and any set of unsafe nodes
$\hat{U}\subseteq{V^U-\{s,t\}}$, the graph $H-\hat{U}$ has
$\min(0,\;r_{st}-|\hat{U}|)$ edge-disjoint $(s,t)$-paths.
For the (uniform~connectivity) $\pNCfgc$ model,
assuming that there is at least one safe node,
we show that there is a 2-approximation algorithm
via a result of Frank \cite[Theorem~4.4]{Frank:dam09}.
}
\end{abstract}

\section{Introduction} \label{sec:intro}
{ 
Adjiashvili, Hommelsheim \& M\"uhlenthaler \cite{AHM21,AHM20}
introduced the model of Flexible Graph Connectivity that we denote by $\fgc$.
Recently, Boyd et~al.\ \cite{BCHI21} introduced a generalization of $\fgc$.
Let $p\geq 1$ and $q\geq 0$ be integers.
In an instance of the $(p,q)$-Flexible Graph Connectivity problem, denoted $\pqfgc$,
we have an undirected connected graph $G = (V,E)$, a partition
of $E$ into a set of safe edges $\safe$ and a set of unsafe edges
$\unsafe$, and nonnegative costs $c\in\Rp^E$ on the edges.
A subset $F \subseteq E$ of edges is feasible for the $\pqfgc$ problem if
for any set $F'\subseteq\unsafe$ with $|F'|\leq q$,
the subgraph $(V, F \setminus F')$ is $p$-edge connected.
The algorithmic goal is to find a feasible solution $F$ that
minimizes $c(F) = \sum_{e \in F} c_e$.

We present approximation algorithms for network design
problems in some models related to the $\pqfgc$ model.

{
First, we present a straight-forward
approximation algorithm for the Flexible Steiner tree problem, denoted $\fst$,
that achieves approximation ratio~$\approx{2.9}$ via recent results
of Ravi, Zhang \& Zlatin \cite{RZZ22}.

Next, we introduce the $\genfgc{\{0,1,\dots,p\}}{\{0,\dots,q\}}$
model, where a required level of edge-connectivity $p_{ij} \in
\{0,1,\dots,p\}$ and a fault-tolerance level $q_{ij} \in \{0,\dots,q\}$
is specified for every pair of nodes $\{i,j\}$.
The goal is to find a subgraph $H=(V,F)$ of minimum cost such that
for any pair of nodes $\{i,j\}$ and any set of at most $q_{ij}$
unsafe edges $F'\subseteq F$, the graph $H - F'$ has $p_{ij}$
edge-disjoint $(i,j)$-paths.
When $q=1$ (i.e., when each $q_{ij}\in\{0,1\}$), we present a
$2(p+1)$-approximation algorithm.
Additionally, when $p=1$ (i.e., when each $p_{ij}\in\{0,1\}$), we
present a $2(q+1)$-approximation algorithm.

Finally, we introduce the \textit{Node-Connectivity} Flexible Graph
Connectivity problem, denoted $\NCfgc$; in this model, the set of
nodes is partitioned into a set of safe nodes (that never fail) and
a set of unsafe nodes (there is no partition of the edge-set);
see Section~\ref{sec:NCfgc} for details.
We observe that there is a simple reduction from the $\NCfgc$ model to
the well-known $\NCsndp$ model;
the latter model has been studied for decades, see \cite{KN:survey2007,Nutov:survey2018}.
Nevertheless, for the (uniform~connectivity) $\pNCfgc$ model,
assuming that there is at least one safe node,
we show that there is a 2-approximation algorithm
via a result of Frank \cite[Theorem~4.4]{Frank:dam09}.
In contrast, for the well-known special case of (uniform~connectivity) $\pNCsndp$,
even with the assumption of $|V(G)|\geq{p^3}$,
the best approximation ratio known is $4+\epsilon$ due to Nutov \cite{Nutov:jcss22}.
}

Other models and results pertaining to Flexible Graph Connectivity
are presented by Adjiashvili, Hommelsheim, M\"uhlenthaler \& Schaudt
\cite{AHMS20} and by Chekuri \& Jain \cite{CJ22a}.
}
\section{Preliminaries} \label{sec:prelims}
{
This section has definitions and preliminary results.
Our notation and terms are consistent with \cite{Diestel,Schrijver},
and readers are referred to these texts for further information.

For a positive integer $k$, we use $[k]$ to denote the set $\{1,\dots,k\}$.

Let $G=(V,E)$ be a (loop-free) multi-graph with non-negative costs $c\in\Rp^{E}$ on the edges.
We take $G$ to be the input graph, and we use $n$ to denote $|V(G)|$.
For a set of edges $F\subseteq E(G)$, $c(F):=\sum_{e\in F}c(e)$,
and for a subgraph $G'$ of $G$, $c(G'):=\sum_{e\in E(G')}c(e)$.
For a graph $H$ and a set of nodes $S\subseteq V(H)$,
$\delta_H(S)$ denotes the set of edges that have one end~node in
$S$ and one end~node in $V(H) \setminus S$;
moreover,
$H[S]$ denotes the subgraph of $H$ induced by $S$, and
$H-S$  denotes the subgraph of $H$ induced by $V(H) \setminus S$.
For a graph $H$ and a set of edges $F\subseteq E(H)$,
$H-F$ denotes the graph $(V(H),~E(H) \setminus F)$.
We may use relaxed notation for singleton sets, e.g.,
we may use $\delta_H(v)$ instead of $\delta_H(\{v\})$, etc.



A multi-graph $H$ is called $k$-edge connected if $|V(H)|\ge2$ and for
every $F\subseteq E(H)$ of size $<k$, $H-F$ is connected.
A multi-graph $H$ is called $k$-node connected if $|V(H)|>k$ and for
every $S\subseteq V(H)$ of size $<k$, $H-S$ is connected.

For any instance $H$, we use $\opt(H)$ to denote the minimum cost of a feasible subgraph
(i.e., a subgraph that satisfies the requirements of the problem).
When there is no danger of ambiguity, we use $\opt$ rather than $\opt(H)$.
}
\section{Extensions of the $\pqfgc$ model with edge-connectivity requirements} \label{sec:fgcextensions}
{
In this section, we present approximation algorithms and reductions
for some extensions of the $\pqfgc$ model with non-uniform connectivity requirements.

First, in Section~\ref{sec:alg-fst}, we present a straight-forward
approximation algorithm for $\fst$
that achieves approximation ratio~$\approx{2.9}$ via recent results
of Ravi, Zhang \& Zlatin \cite{RZZ22}.

Next, in Section~\ref{sec:nonuniformfgc}, we introduce the
$\genfgc{\{0,1,\dots,p\}}{\{0,\dots,q\}}$ model, and we present
approximation algorithms when $q=1$ (i.e., when each $q_{ij}\in\{0,1\}$)
or
when $p=1$ (i.e., when each $p_{ij}\in\{0,1\}$).

\subsection{Simple approximation algorithm for $\fst$} \label{sec:alg-fst}
{
This section presents and analyzes a simple (and obvious) two-stage
approximation algorithm for $\fst$ 
that achieves approximation ratio~$\approx{2.9}$ via recent results
of Ravi, Zhang \& Zlatin \cite{RZZ22}.

The first-stage algorithm applies the best-known approximation algorithm for
the Steiner tree problem, due to \cite{BGRS10,BGRS13},
to the input, and makes no distinction between safe edges and unsafe edges.
Let $H_1=(V_1,F_1)$ denote the Steiner tree found by this algorithm;
clearly, $c(F_1) \leq 1.4 \,\opt$.

The second-stage algorithm applies Jain's iterative rounding 2-approximation algorithm,
\cite{Jain01}, to the following instance of SNDP (i.e., Survivable Network Design Problem):
Let $G_2=G/(F_1\cap\safe)$ be the graph obtained from $G$ by
contracting the safe edges of $F_1$, and fix the edge-costs $c''$
of $G_2$ by fixing $c''_e=0$ for each edge $e\in{F_1\cap\unsafe}$
and fixing $c''_e=c_e$ for the other edges $e$; moreover, let the
set of terminals be $\term_2=\term$ if $V(F_1\cap\safe)$ has no
terminals, otherwise, let
$\term_2=\hat{\term}\cup(\term\setminus{V(F_1\cap\safe)})$, where
$\hat{\term}$ denotes the set of contracted nodes $\hat{u}$ of $G_2$
such that the set of edges of $G$ whose contraction results in
$\hat{u}$ is incident to a terminal; the connectivity requirements
are two for each pair of terminals (and zero for all other node pairs).
(In this instance, there is no distinction between safe edges and unsafe edges.)
Let $H_2=(V_2,F_2)$ denote the $\twoecs$ found by this algorithm.
We claim that (i)~$c(F_2) \leq 2\,\opt$, and, moreover,
(ii)~$(V_1\cup{V_2},\,F_1\cup{F_2})$
is a feasible subgraph of the $\fst$ instance.
To prove claim~(i), consider the subgraph $H_0$ formed by the edge-set
$(F^* \cup F_1)/(F_1\cap\safe)$,
where $F^*$ denotes the edge-set of an optimal subgraph of the $\fst$ instance;
observe that for any edge $e$, $H_0-e$ contains a Steiner tree on~$\term_2$,
that is, $H_0$ has two edge-disjoint paths between each pair of nodes of $\term_2$
\big(in more detail, if $e$ is a safe edge, then
$F_1/(F_1\cap\safe)=F_1\cap\unsafe$ contains a Steiner tree
on~$\term_2$, and if $e$ is an unsafe edge, then $F^*\setminus\{e\}$
contains a Steiner tree on~$\term$, so $F^*/(F_1\cap\safe)$ contains
a Steiner tree on~$\term_2$\big);
clearly, $c''(H_0) \leq \opt$.
Claim~(ii) follows easily from the fact that $H_2$ has two edge-disjoint
paths between each pair of nodes of $\term_2$.

We remark that the approximation guarantee of the above algorithm can be improved by using better algorithms for the second-stage. In particular the second-stage algorithm is equivalent to the problem of augmenting a Steiner tree. Recently, Ravi, Zhang \& Zlatin \cite{RZZ22} provided a $(1.5+\epsilon)$ approximation algorithm for this problem. Hence, the two-stage approximation algorithm for FST achieves approximation ratio $\approx{2.9}$.
}
\subsection{$\fgc$ with non-uniform edge-connectivity requirements and fault-tolerance} \label{sec:nonuniformfgc}
{
In this section, we study the $\genfgc{\{0,1,\dots,p\}}{\{0,\dots,q\}}$~model.
This model is a generalization of the $\pqfgc$~model
in \cite{BCHI21} with non-uniform edge-connectivity requirements.
The input consists of an undirected graph $G = (V,E)$ with non-negative
costs on the edges $c\in\Rp^E$, and a partition of the edge-set $E$
into a set $\safe$ of \textit{safe edges} and a set $\unsafe$ of
\textit{unsafe edges}; additionally, for every pair of nodes $\{i,j\}$,
a required level of edge-connectivity $p_{ij} \in \{0,1,\dots,p\}$ and a
fault-tolerance $q_{ij} \in \{0,\dots,q\}$ is specified.
The goal is to find a subgraph $H = (V,F)$ of minimum cost such
that for any pair of nodes $(i,j)$ and any set of at most $q_{ij}$
unsafe edges $F'\subseteq F$, the graph $H\setminus F'$ has at least
$p_{ij}$ edge-disjoint $(i,j)$-paths.
We present a $2(p+1)$-approximation algorithm 
when $q=1$ (i.e., when each $q_{ij}\in\{0,1\}$).
We also provide a $2(q+1)$-approximation algorithm
when $p=1$ (i.e., when each $p_{ij}\in\{0,1\}$).
The main idea is to formulate the problems as a capacitated network
design problem ($\capndp$), and then apply Jain's iterative rounding
method, \cite{Jain01}, to solve the resulting $\capndp$~instance.

\subsubsection{$\genfgc{\{0,1,\dots,p\}}{\{0,\dots,q\}}$ with $q=1$}
{
First, we consider the $\genfgc{\{0,1,\dots,p\}}{\{0,\dots,q\}}$~model 
when $q=1$ (i.e., when each $q_{ij}\in\{0,1\}$).
We formulate a corresponding instance of $\capndp$ where the capacity of an edge
$e$ is equal to $p+1$ if edge $e$ is safe and is equal to $p$
otherwise (thus, $u_e=p$ if $e$ is unsafe and $u_e=p+1$ if $e$ is
safe); moreover, in the $\capndp$ instance, fix the demand between
a pair of nodes $\{i,j\}$ to be $D_{ij} = (p+q_{ij})p_{ij}$. This gives
rise to the following integer program.

\begin{align*}
    \min\;\;\;\; &\sum_{e\in E}c_e x_e\\
     s.t.\;\;\;\; &\sum_{e\in \delta(S)} u_e x_e \geq \max_{(i,j): i\in S, j\not\in S} D_{ij} && \forall \emptyset \neq S \neq V\\
     & x_e \in \{0,1\} && \forall e\in E
\end{align*}

\begin{lemma}
A feasible solution $H=(V,F)$ of the $\genfgc{\{0,1,\dots,p\}}{\{0,1\}}$
problem corresponds to a feasible solution of the above integer program.
\end{lemma}

\begin{proof}
A feasible solution $H=(V,F)$ is characterized by the following
property: For any $(i,j)$-cut $\delta_H(S)$ of $H$, the cut contains
either $p_{ij}$ safe edges or at least $p_{ij}+q_{ij}$ edges in
total.  We claim that feasible solutions to the above integer program
are also characterized by the same property. Indeed if the cut
$\delta_H(S)$ has at least $p_{ij}+q_{ij}$ edges in total, then the
capacity across this cut is at least $(p_{ij}+q_{ij})p \geq p_{ij}p
+ q_{ij}p_{ij} = D_{ij}$.  If the cut $\delta_H(S)$ has fewer than
$p_{ij}+q_{ij}$ edges and exactly $p_{ij}$ edges, then $q_{ij}=1$.
We argue that in this case all edges in this cut must be safe edges,
otherwise, the capacity across this cut will be strictly less than
$p_{ij}(p+1) = D_{ij}$. Finally, if the cut $\delta_H(S)$
has fewer than $p_{ij}$ edges, then the capacity across this cut
is at most $(p_{ij}-1)(p+1) < D_{ij}$.
\end{proof}

To solve (approximately) the above instance of $\capndp$, we replace
each edge $e\in E$ by $u_e$ parallel edges $e'_1,\ldots,e'_{u_e}$
of capacity one and cost $c_e$ each, obtaining an instance of the
survivable network design problem ($\sndp$). Suppose that $F^*$ is
an optimal solution to the $\capndp$ instance.  Then taking all the
edges $\{e'_1,\ldots,e'_{u_e}\}_{e\in F^*}$ gives us a feasible
solution to the $\sndp$ instance. The cost of this solution is at
most $(p+1)c(F^*)$. Using Jain's iterative rounding method,
\cite{Jain01}, we can obtain a solution $F'$ to the $\sndp$ instance
with cost at most $2(p+1)c(F^*)$.  Finally, by picking edges $e\in
E$ if $e_i \in F'$ for any $i=1,\ldots,u_e$, we obtain a feasible
solution to the $\capndp$ instance with cost at most $2(p+1)c(F^*)$.
}
\subsubsection{$\genfgc{\{0,1,\dots,p\}}{\{0,\dots,q\}}$ with $p=1$}
{
Next, we consider the $\genfgc{\{0,1,\dots,p\}}{\{0,\dots,q\}}$~model 
when $p=1$ (i.e., when each $p_{ij}\in\{0,1\}$).
We formulate a corresponding instance of $\capndp$ where the capacity of an edge
$e$ is equal to $q+1$ if edge $e$ is safe and is equal to $1$
otherwise (thus, $u_e=1$ if $e$ is unsafe and $u_e=q+1$ if $e$ is
safe); moreover, in the $\capndp$ instance, fix the demand between
a pair of nodes $\{i,j\}$ to be $D_{ij} = (q_{ij}+1)p_{ij}$. This gives
rise to the following integer program.

\begin{align*}
    \min\;\;\;\; &\sum_{e\in E}c_e x_e\\
     s.t.\;\;\;\; &\sum_{e\in \delta(S)} u_e x_e \geq \max_{(i,j): i\in S, j\not\in S} D_{ij} && \forall \emptyset \neq S \neq V\\
     & x_e \in \{0,1\} && \forall e\in E
\end{align*}

\begin{lemma}
A feasible solution $H=(V,F)$ of the $\genfgc{\{0,1\}}{\{0,\dots,q\}}$
problem corresponds to a feasible solution of the above integer program.
\end{lemma}

\begin{proof}
A feasible solution $H = (V,F)$ is characterized by the following
property: For any $(i,j)$-cut $\delta_H(S)$ of $H$, the cut contains
either $p_{ij}\in \{0,1\}$ safe edges or at least $p_{ij}+q_{ij}$ edges in
total. We claim that feasible solutions to the above integer program
are also characterized by the same property. Indeed if the cut $\delta_H(S)$
has at least $p_{ij}+q_{ij}$ edges in total, then the capacity
across this cut is at least $p_{ij}(q_{ij}+1) = D_{ij}$. If
the cut $\delta_H(S)$ has at least $p_{ij}$ safe edges, then the capacity
across this cut is at least $(q+1)p_{ij} \geq (q_{ij}+1)p_{ij} =
D_{ij}$. Finally, if the cut $\delta_H(S)$ contains fewer than
$p_{ij}+q_{ij}$ edges in total and fewer than $p_{ij}$ safe
edges, then we must have $p_{ij} = 1$, so then, the cut contains no safe edges.
Hence, the capacity across this cut is at most $q_{ij} < D_{ij}$.
\end{proof}

We have already seen how to solve (approximately) the above instance
of Cap-NDP and thus we can obtain a feasible solution with cost at
most $2(q+1)c(F^*)$, where $F^*$ is an optimal solution.
}
}
}
\section{The $\NCfgc$ model} \label{sec:NCfgc}
{
In this section, we introduce the
\textit{Node-Connectivity} Flexible Graph Connectivity problem, denoted $\NCfgc$.
We observe that there is a simple reduction from the $\NCfgc$ model to
the well-known $\NCsndp$ model;
the latter model has been studied for decades, see \cite{KN:survey2007,Nutov:survey2018}.
Nevertheless, for the (uniform~connectivity) $\pNCfgc$ model,
assuming that there is at least one safe node,
we show that there is a 2-approximation algorithm
via a result of Frank \cite[Theorem~4.4]{Frank:dam09}.
In contrast, for the well-known special case of (uniform~connectivity) $\pNCsndp$,
even with the assumption of $|V(G)|\geq{p^3}$,
the best approximation ratio known is $4+\epsilon$ due to Nutov \cite{Nutov:jcss22}.

In an instance of the $\NCfgc$ problem, we have an undirected connected graph
$G = (V,E)$, a partition of $V$ into a set of safe nodes $V^S$ and a
set of unsafe nodes $V^U$, and non-negative costs $c\in\Rp^E$ on the
edges; moreover, for every pair of nodes $\{s,t\}$, a required level of
connectivity $r_{st}\in\Zintp$ is specified.
The goal is to find a subgraph $H=(V,F)$ of minimum cost such that for
any pair of nodes $\{s,t\}$, and any set of unsafe nodes
$\hat{U}\subseteq{V^U-\{s,t\}}$, the graph $H-\hat{U}$ has
$\min(0,\;r_{st}-|\hat{U}|)$ edge-disjoint $(s,t)$-paths.
The goal can be re-stated using the notion of $q$-connectivity,
see Nutov \cite{Nutov:survey2018}.
Given node~capacities $q\in\Zintp^V$, the \textit{$q$-connectivity}
of a pair of nodes $\{s,t\}$ of a graph $H$, denoted $\lambda^q_H(s,t)$,
is the maximum number of pairwise edge disjoint $(s,t)$-paths such
that each node $v\in{V-\{s,t\}}$ is in $\leq q_v$ of these paths.
We fix $q_v=\infty$ for each safe node $v$, and
we fix $q_u=1$ for each unsafe node $u$.
Then, the goal is to find a minimum cost subgraph $H=(V,F)$
such that $\lambda^q_H(s,t)\geq{r_{st}}$ for each pair of nodes $\{s,t\}$.
For some special cases of the $q$-connectivity model,
the known approximation algorithms for the well-known $\NCsndp$
model extend to the $q$-connectivity model with the same approximation
ratio, see \cite{Nutov:survey2018} and \cite{KN:survey2007}.
To see this directly for the $\NCfgc$ model (with
$q\in\{1,\infty\}^V$), one can ``inflate'' each safe node $v$ of $G$ to
a complete graph $K_v$ on $\deg_G(v)$ nodes with edges of cost zero, and
replace the edges incident to $v$ (in $G$) by edges incident to distinct nodes of $K_v$
(in the inflated graph) while preserving the edge costs;
thus, any $(s,t)$-path of $G$ maps to a $(K_s,K_t)$-path of the inflated graph and vice versa.

Let $\pNCfgc$ denote the special case of the $\NCfgc$ model with a
uniform~connectivity requirement of $r_{st}=p$ for every pair of nodes $\{s,t\}$.
We apply a result of Frank, \cite[Theorem~4.4]{Frank:dam09}, to present
a 2-approximation algorithm for the $\pNCfgc$ model, assuming that
there is at least one safe node.

\begin{proposition} \label{propos:pNCfgc}
Given an instance $G=(V,E),\;c\in\Rp^E,\;p$ of $\pNCfgc$
such that $V^S\not=\emptyset$,
there is an (polynomial-time) algorithm that finds a feasible subgraph
of cost $\leq2\cdot\opt$.
\end{proposition}

\begin{proof}
We construct a digraph $D$ by replacing each edge $e$ of $G$ by a pair
of anti-parallel arcs that each have cost $c_e$.
We pick the root $s_0$ to be any safe node of $D$.
Then we assign node capacities to the nodes of $D$:
we fix $q_v=p$ for each safe node $v$, and
we fix $q_u=1$ for each unsafe node $u$.
Consider the (directed) rooted $q$-connectivity problem for
$D,\,q,\,c,\,s_0$:  the goal is to find a minimum-cost subgraph ${\dH}$ of
$D$ such that $\dlambda^q_{\dH}(s_0,t)\geq{p}$ for each node
$t\in{V(D)-{s_0}}$, where $\dlambda^q_{\dH}(s_0,t)$ denotes the maximum
number of $(s_0,t)$-dipaths in ${\dH}$ such that each node
$v\in{V-\{s_0,t\}}$ is in at most $q(v)$ of these dipaths.
Frank, \cite[Theorem~4.4]{Frank:dam09}, presents a reduction from the
(directed) rooted $q$-connectivity problem to weighted matroid
intersection. Thus, we can find a minimum-cost rooted $q$-connected
subgraph ${\dH}$ of $D$.

Finally, we return the subgraph $H$ of $G$ that corresponds to ${\dH}$,
where $H$ has an edge $ij$ iff ${\dH}$ contains one of the arcs $(i,j)$
or $(j,i)$. We claim that $H$ is a feasible subgraph of the instance of
$\pNCfgc$, that is, $\lambda^q_H(s,t)\geq p$ for every pair of nodes
$\{s,t\}$, where $q$ is as above.
To verify this, consider a pair of nodes $\{s,t\}$ and any set $\hat{U}
\subseteq{V^U-\{s,t\}}$ of size $\leq{p-1}$.  Observe that
${\dH}-\hat{U}$ has $p-|\hat{U}|$ arc disjoint $(s_0,t)$-dipaths (such
that these dipaths contain at most one arc from each pair of
anti-parallel arcs); hence, $H-\hat{U}$ has $p-|\hat{U}|$ edge disjoint
$(s_0,t)$-paths. Similarly, it follows that $H-\hat{U}$ has
$p-|\hat{U}|$ edge disjoint $(s_0,s)$-paths. Hence, $H-\hat{U}$ has
$p-|\hat{U}|$ edge disjoint $(s,t)$-paths.
This proves our claim that $H$ is a feasible subgraph of the instance
of $\pNCfgc$.

Observe that $H$ has cost $\leq2\cdot\opt$, because the optimal
subgraph of the instance of $\pNCfgc$ corresponds to a subgraph of $D$
of cost $2\cdot\opt$ that is feasible for the above (directed) rooted
$q$-connectivity problem.
\end{proof}
}

\vfill
\newpage
\bibliographystyle{plainurl}
\bibliography{bcgi-extensions-ref}
\end{document}